\documentclass[prb,twocolumn,superscriptaddress,fixfloat]{revtex4}
\usepackage{amsmath}
\usepackage{amssymb}
\usepackage{graphicx}
\usepackage{color}  
\begin{document}

\title{Dispersive Spin Fluctuations in the near optimally-doped superconductor Ba(Fe$_{1-x}$Co$_{x}$)$_2$As$_2$ ($x$=0.065)}

\author{C. Lester}
\affiliation{H.H. Wills Physics Laboratory, University of Bristol, Tyndall Ave., Bristol, BS8 1TL, United Kingdom}

\author{Jiun-Haw Chu}
\affiliation{Geballe Laboratory for Advanced Materials and Department of Applied Physics, Stanford University, Stanford, CA 94305}
\affiliation{Stanford Institute for Materials and Energy Sciences, SLAC National Accelerator Laboratory, 2575 Sand Hill Road, Menlo Park, CA 94025}

\author{J. G. Analytis}
\affiliation{Geballe Laboratory for Advanced Materials and Department of Applied Physics, Stanford University, Stanford, CA 94305}
\affiliation{Stanford Institute for Materials and Energy Sciences, SLAC National Accelerator Laboratory, 2575 Sand Hill Road, Menlo Park, CA 94025}

\author{T. G. Perring}
\affiliation{ISIS Facility, STFC Rutherford Appleton Laboratory, Chilton, Didcot, OX11 0QX, United Kingdom}

\author{I. R. Fisher}
\affiliation{Geballe Laboratory for Advanced Materials and Department of Applied Physics, Stanford University, Stanford, CA 94305}
\affiliation{Stanford Institute for Materials and Energy Sciences, SLAC National Accelerator Laboratory, 2575 Sand Hill Road, Menlo Park, CA 94025}

\author{S.M. Hayden}
\affiliation{H.H. Wills Physics Laboratory, University of Bristol, Tyndall Ave., Bristol, BS8 1TL, United Kingdom}

\begin{abstract}
Inelastic neutron scattering is used to probe the collective spin excitations of the near optimally-doped  superconductor Ba(Fe$_{1-x}$Co$_{x}$)$_2$As$_2$ ($x$=0.065).  Previous measurements on the antiferromagnetically ordered parents of this material show a strongly anisotropic spin-wave velocity. Here we measure the magnetic excitations up to 80 meV and show that a similar anisotropy persists for superconducting compositions. The dispersive mode measured here connects directly with the spin resonance previously observed in this compound.  When placed on an absolute scale, our measurements show that the local- or wavevector- integrated susceptibility is larger in magnitude than that of the ordered parents over the energy range probed.
\end{abstract}


\maketitle

\section{Introduction}

The recently discovered ferropnictide superconductors \cite{Kamihara2008a} show critical temperatures exceeding 50~K. They are interesting materials, both in their own right and because they provide valuable insights into other classes of superconductor.  Theoretical calculations \cite{Boeri2008a} suggest that the electron-phonon coupling is too weak in the ferropnictides to produce the high transition temperatures which are observed experimentally. The ferropnictides have some similarities to the cuprates in that they are quasi two-dimensional and have antiferromagnetic parent compounds.  Superconductivity in the ferropnictides can be induced \cite{Rotter2008a,Sefat2008a} from their antiferromagnetic parents by various means: electron or hole doping via chemical substitution; isovalent substitution of the iron or arsenic or by the application of pressure. In the widely-studied AFe$_{2}$As$_{2}$ (A=Ca,Sr,Ba) family (known as the ``122'' family)  \cite{Rotter2008a,Sefat2008a,Ni2008a,Chu2009a,Lester2009a}, this can be via chemical substitution at the Fe or As sites or the application of pressure.   For example, in BaFe$_{2}$As$_{2}$, we obtain superconductivity through electron doping by Co or Ni substitution at the Fe site; by hole doping with K at the Ba site; or by isovalent substitution of Fe by Ru or As by P.

The proximity of superconductivity to antiferromagnetism suggests that the pairing mechanism in the doped ferropnictides is related to the spin degrees of freedom. Theories based on various models of the magnetic excitations have been proposed \cite{Mazin2008a,Barzykin2008a,Korshunov2008a,Cvetkovic2009a,Chubukov2008a,Graser2009a}.
In order to take such theories forward, the collective magnetic excitations in the superconducting region of the phase diagram need to be characterized and the underlying interactions understood.
Thus, in this paper, we report an inelastic neutron scattering (INS) study of the magnetic response, up to 80~meV, of a near optimally-doped superconducting ($T_c$=23~K) composition Ba(Fe$_{0.935}$Co$_{0.065}$)$_2$As$_2$.  This composition appears not to be magnetically ordered. We compare our results with similar measurements on the antiferromagnetic parent compounds \cite{Ewings2008a,Matan2009a,Diallo2009a,Zhao2009a} of the 122 series and find that (i) the magnetic excitations are stronger than in the parent compounds and (ii) the anisotropy in the spin-wave velocity observed in the parent compounds persists in this superconductor.

\section{Experimental Details}

Single crystals of Ba(Fe$_{0.935}$Co$_{0.065}$)$_2$As$_2$ were grown by a self-flux method \cite{Chu2009a}. 15 crystals were co-aligned on a thin Al plate using x-rays and neutron diffraction. Our neutron measurements were made on an mosaic of total mass 0.3~g. Resistivity and magnetization measurements identified the superconducting transition temperature $T_c$(onset) = 23~K. Elastic neutron scattering revealed no evidence of magnetic order at this doping level at temperatures down to 2~K. We used the MAPS instrument at the ISIS spallation source. MAPS is a low-background direct-geometry time-of-flight chopper spectrometer with position sensitive detectors \cite{MAPS_ref}. A pulse of spallation neutrons spread over a time of about 5-10~$\mu$s is produced when a pulse of protons hits a Ta target adjacent to a water moderator. Neutrons with the required energy ($E_i$=60, 80, or 140~meV in the present experiment) are then selected by an appropriately phased Fermi chopper (rotating at 100~Hz in the present experiment). The Fermi chopper is 10~m from the neutron source and opens for about 45~$\mu$s. The neutrons then scatter from the sample (12~m from source) and are detected in position sensitive $^{3}$He detectors at 6~m from the sample. The detection time of the neutron is used to determine its energy transfer.   Data are averaged of a range of energies to improve the experimental statistics.  The energy ranges of integration are given when we quote neutron energy transfers in the text.  Data were placed on an absolute scale (barn sr$^{-1}$ f.u.$^{-1}$) by comparing the count rate with that from a plate of vanadium.

\begin{figure}
\begin{center}
\includegraphics[width=0.95\linewidth]{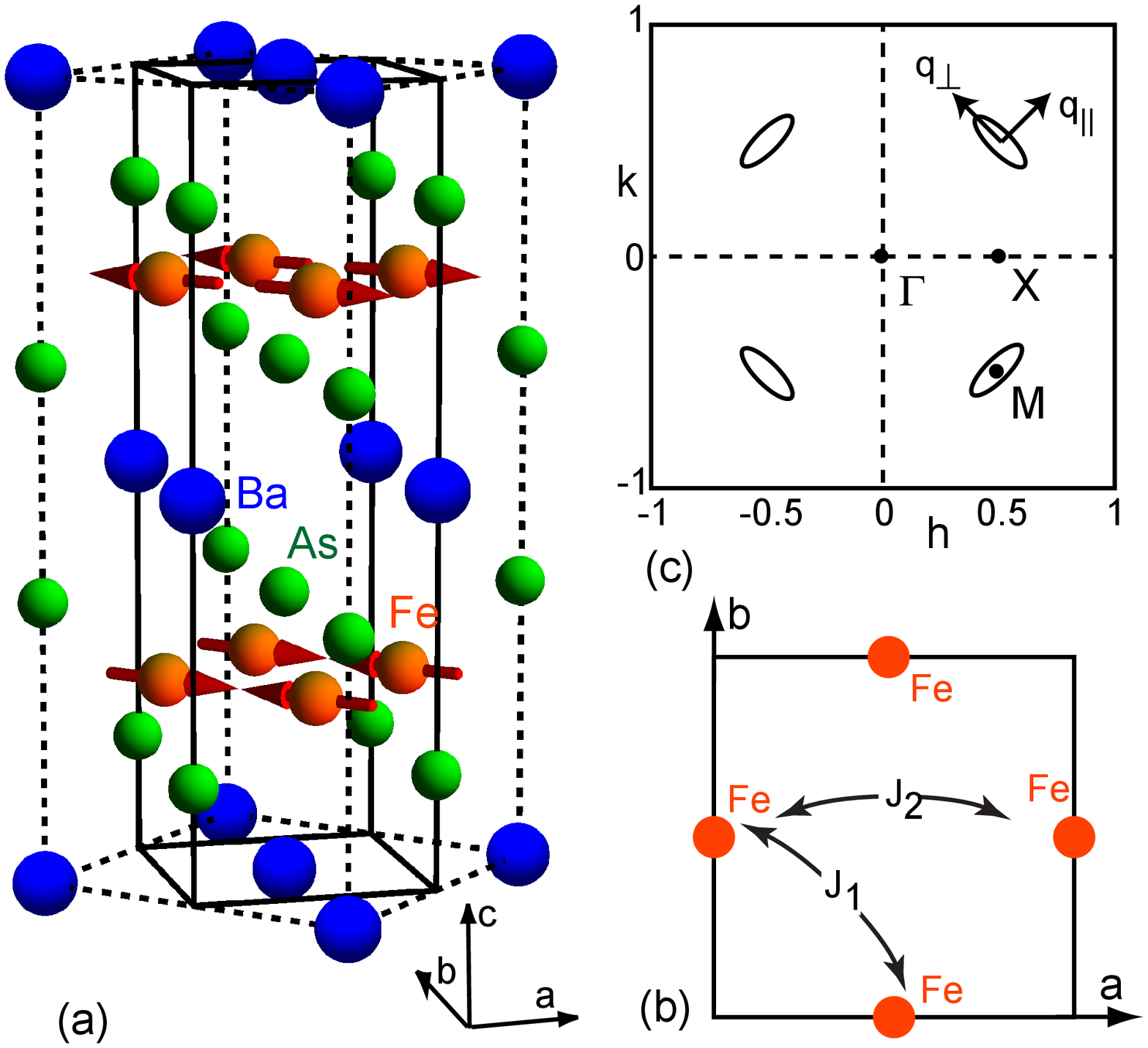}
\end{center}
\caption{(color online) (a) The antiferromagnetic structure of the parent compound BaFe$_2$As$_2$.   Axes refer to the tetragonal notation used throughout the paper. Solid lines denote the tetragonal unit cell. (b) Exchange couplings between Fe atoms referred to in the text.  (c) Schematic illustration of reciprocal space (of a 2D square lattice). The ellipses represent the anisotropic excitations reported in this paper. The wavevectors are used to define the spin wave velocities.}
\label{Fig:unit_cell}
\end{figure}

The magnetic cross section of an isotropic paramagnet is given by
\begin{equation}
\label{Eq:cross_sect} \frac{d^2\sigma}{d\Omega \, dE} = \frac{2(\gamma
r_{\text{e}})^2}{\pi g^{2} \mu^{2}_{\rm B}} \frac{k_f}{k_i} \left| F({\bf Q})\right|^2
\frac{\chi^{\prime\prime}({\bf q},\omega)}{1-\exp(-\hbar\omega/kT)},
\end{equation}
where $(\gamma r_{\text{e}})^2$=0.2905 barn sr$^{-1}$, ${\bf k}_{i}$ and ${\bf k}_{f}$ are the incident and final neutron wavevectors and $|F({\bf Q})|^2$ is the isotropic magnetic form
factor for a Fe$^{2+}$ orbital.  We use Eq.~\ref{Eq:cross_sect} to convert the measured cross section to the energy- and momentum-dependent susceptibility \cite{chi_comment} $\chi^{\prime\prime}(\textbf{q},\omega)$.
Ba(Fe$_{0.935}$Co$_{0.065}$)$_2$As$_2$ has the tetragonal crystal
structure shown in Fig.~\ref{Fig:unit_cell}(a) with lattice
parameters $a$=3.955~\AA\ and $c$=12.95~\AA. We use the reciprocal
space notation to label wavevectors
$\mathbf{Q}=h\mathbf{a}^{\star}+k\mathbf{b}^{\star}+l\mathbf{c}^{\star}$.
In this notation, the antiferromagnetic ordering wavevector of
BaFe$_{2}$As$_2$ is (1/2,1/2,1).  The data reported in this paper were collected with $\mathbf{c}^{\star}$ parallel to $\mathbf{k}_i$.  Under these conditions, there is a coupling of $l$ and the $\omega$.  We give the $l$ value corresponding to each $\omega$ in the figures and captions.

\section{Results}
\begin{figure*}
\begin{center}
\includegraphics[width=0.8\linewidth]{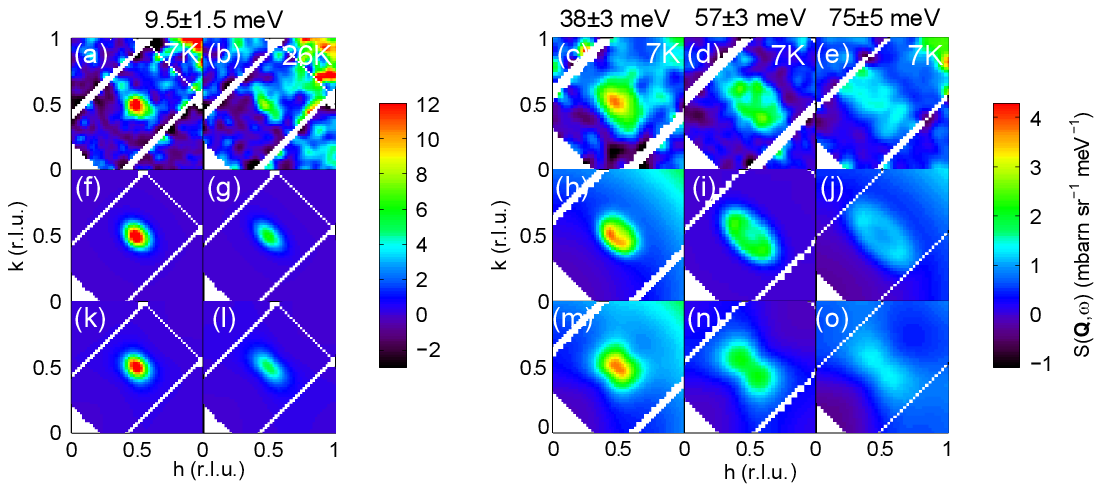}
\end{center}
\caption{(color online) (a)-(e) Constant-energy slices through the magnetic
excitations in Ba(Fe$_{0.935}$Co$_{0.065}$)$_2$As$_2$ as observed on MAPS. (f)-(j)  Fits to a phenomenological spin-wave cross-section (Eqs.~\ref{Eq:DHO}-\ref{Eq:dispersion}). (k)-(o) Fits to a phenomenological Sato-Maki cross sections (Eq.~\ref{Eq:Sato_Maki}) [e.g. for (n) $\kappa=0.21 \pm 0.09$, $\delta=0.6 \pm 0.2$, $\lambda=5 \pm 2$].  The incident energies used were 60 meV [(a)-(b)] and 140 meV [(c)-(e)] and the corresponding $l$ values were $l$=1 [(a)-(b)], 2.5 (c), 4(d) and 5.5 (e). The $E_i$=140 meV data were collected using a proton beam current of 175~$\mu$A for 80 hours.  A constant background has been subtracted from each plot.} \label{Fig:slices}
\end{figure*}

\begin{figure}
\begin{center}
\includegraphics[width=0.85\linewidth]{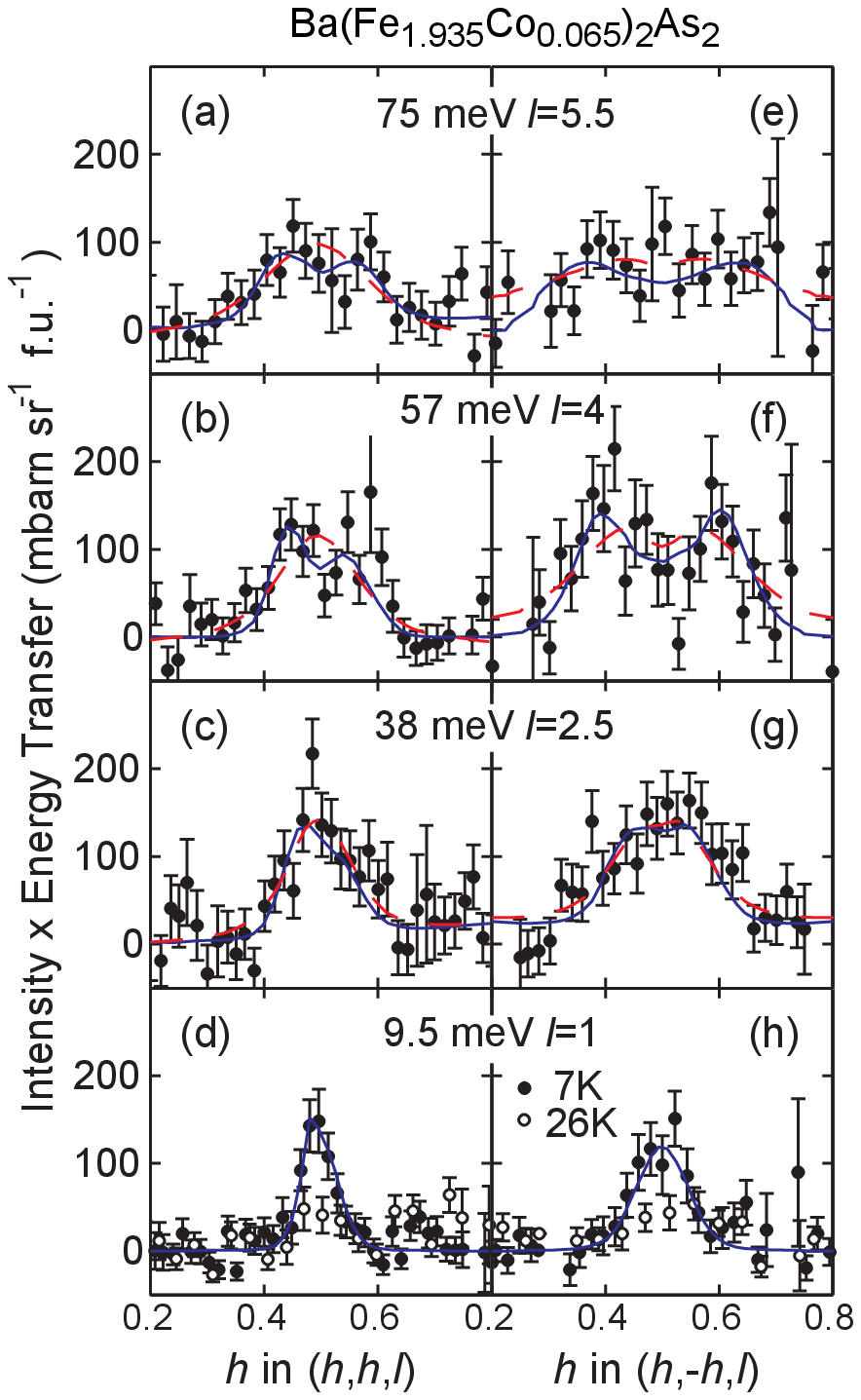}
\end{center}
\caption{(Color online) Constant $\hbar\omega$ and $l$ cuts along $(1/2-h,1/2+h,l)$ and $(1/2-h,1/2+h,l)$ (see Fig.~\ref{Fig:unit_cell}) for various values of $\hbar\omega$. The blue solid line a fit to the SW cross-section Eqs.~\ref{Eq:DHO}-\ref{Eq:dispersion} and the red dotted line to the Sato-Maki model (Eq.~\ref{Eq:Sato_Maki}). Where only one line is drawn, the fits are indistinguishable.}
\label{Fig:cuts}
\end{figure}

Figure~\ref{Fig:slices} shows typical $\mathbf{q}$-dependent images of $S(\textbf{q},\omega)=(k_i/k_f) d^2\sigma /d\Omega dE$ in the $(h,k)$ plane for various energy transfers obtained from the MAPS spectrometer.  At first sight, the data look similar to those obtained over the same energy range on the related antiferromagnetic parent compound CaFe$_{2}$As$_{2}$ \cite{Diallo2009a,Zhao2009a}.  At low energies, $E$=$9.5 \pm 1.5$~meV [Fig.~\ref{Fig:slices}(a)], the magnetic response is strongest at the (1/2,1/2) position. The excitations disperse with increasing energy with the pattern broadening most rapidly along the $(1/2-\xi,1/2+\xi)$ rather than the $(1/2+\xi,1/2+\xi)$ direction, leading to an elliptically shaped response as shown schematically in Fig.~\ref{Fig:unit_cell}(b).
Fig.~\ref{Fig:cuts} shows cuts through the data shown in Fig.~\ref{Fig:slices}, again illustrating the anisotropic dispersion and broadening of the response with increasing energy transfer.

\begin{figure}
\begin{center}
\includegraphics[width=0.7\linewidth]{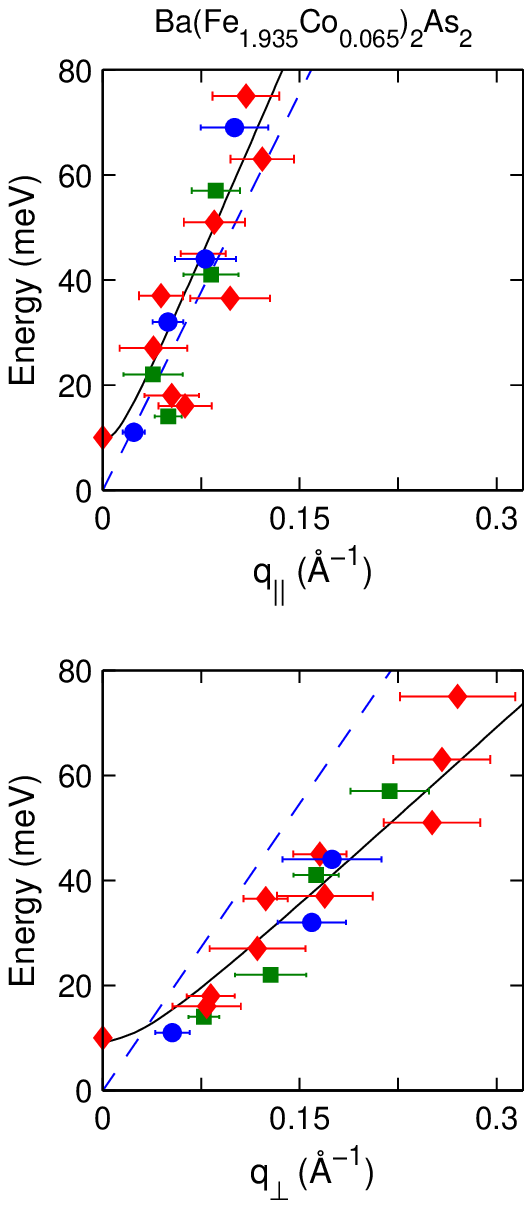}
\end{center}
\caption{(Color online) The dispersion of the magnetic excitations in
Ba(Fe$_{0.935}$Co$_{0.065}$)$_2$As$_2$ for $T=7$~K as determined from fitting
the damped spin-wave cross section
(Eqs.~\ref{Eq:dispersion}--\ref{Eq:DHO}) along q$_{\parallel}$ (a)
and q$_{\perp}$ (b) compared to that found in CaFe$_2$As$_2$ \cite{Zhao2009a} (blue
dashed line) in each case.  Symbols denote $l$ values: {\scriptsize $\blacksquare$} ($l$ even); {\large $\bullet$} ($l$ odd); $\blacklozenge$ ($l$ non-integer).
}
\label{Fig:dispersion}
\end{figure}

\begin{figure}
\begin{center}
\includegraphics[width=0.7\linewidth]{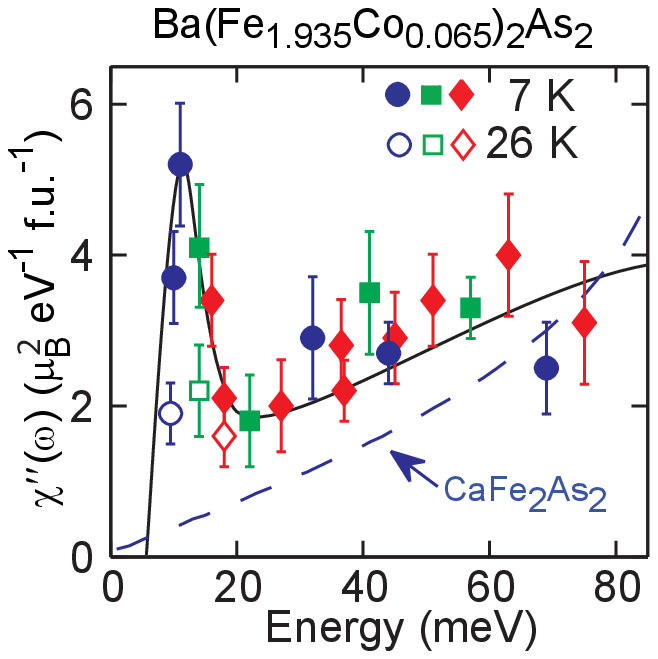}
\end{center}
\caption{(Color online) The local susceptibility
$\chi^{\prime\prime}(\omega)$ at 7~K (closed symbols) and 26~K (open
symbols). Symbols denote $l$ values: {\scriptsize $\blacksquare$} ($l$ even); {\large $\bullet$} ($l$ odd); $\blacklozenge$ ($l$ non-integer). Dashed line shows the local susceptibility for CaFe$_2$As$_2$ \cite{Zhao2009a,Zhao_comment}.}
\label{Fig:chi_local}
\end{figure}

Although Ba(Fe$_{0.935}$Co$_{0.065}$)$_2$As$_2$ is not magnetically ordered, its response is reminiscent of the magnetically ordered 122 parent compounds.  In order to make a more quantitative analysis of our data we fitted a phenomenological spin-wave cross section with a damped-harmonic-oscillator (DHO) line shape in energy. The DHO cross-section is specified by Eqs.~\ref{Eq:DHO}--\ref{Eq:dispersion}:
\begin{equation}
\label{Eq:DHO}
\chi^{\prime\prime}(\mathbf{q},\omega)=\chi^{\prime}(\mathbf{q},0)\frac{2 \omega_{0}^2 \Gamma \omega}{(\omega_{0}^2-\omega^2)^2+(2 \omega \Gamma)^2},
\end{equation}
where
\begin{equation}
\label{Eq:dispersion}
\omega_{0}=\left[ (v_{\parallel}  q_{\parallel})^{2} + (v_{\perp}  q_{\perp})^{2} + \Delta^2 \right]^{1/2}.
\end{equation}
In these equations, $v_{\parallel}$ and $v_{\perp}$ describe the dispersion along $(1/2+\xi,1/2+\xi)$-type and $(1/2-\xi,1/2+\xi)$-type directions respectively (or $q_{\parallel}$ and $q_{\perp}$ in Fig.~\ref{Fig:unit_cell}), $\Gamma$ is the damping parameter and $\Delta$ is a gap (see later).  Eqs.~\ref{Eq:DHO}--\ref{Eq:dispersion} give a good description of the data at all energies.  The $\mathbf{q}$-averaged or local susceptibility $\chi^{\prime\prime}(\omega)$ is defined as:
\begin{equation}
\label{Eq:chi_local}
\chi^{\prime\prime}(\omega)=\frac{\int \chi^{\prime\prime}(\mathbf{q},\omega) \; d\mathbf{q}}{\int d\mathbf{q}},
\end{equation}
where the average is over sufficient Brillouin zones in the extended zone scheme to sample representative wavevectors of $\chi^{\prime\prime}(\mathbf{q},\omega)$. The local quantity gives an indication of the overall strength of the excitations and can be used to compute the corresponding fluctuating moment
 \begin{equation}
\label{Eq:moment_sqr}
\langle \mathbf{m}^2 \rangle = \frac{3 \hbar}{\pi} \int_{-\infty}^{\infty} \frac{\chi^{\prime\prime}(\omega) \; d\omega}{1-\exp(-\hbar \omega/kT)}.
\end{equation}
Figs.~\ref{Fig:dispersion}-\ref{Fig:chi_local} show the dispersion and the energy dependence of $\chi^{\prime\prime}(\omega)$ determined from fits to the data. The different symbols in Figs.~\ref{Fig:dispersion} and \ref{Fig:chi_local} represent different $l$ values for the $c$-axis momentum.  In the present experiment we average $\chi^{\prime\prime}(\mathbf{Q},\omega)$ over $0<h<1$ and $0<k<1$ and $l$ over approximately $\pm 0.25$ to compute the values in Fig.~\ref{Fig:chi_local}.  All points appear to follow the same trend and thus we are unable to discern any evidence for magnetic coupling in the $c$-direction for the energy scale of the present experiment $E \gtrsim 10$~meV.

A second function which has been used to describe the magnetic response in nested Fermi liquids is the modified Lorentzian or Sato-Maki (SM) function \cite{Sato1974a}.
This function has been successfully used to model chromium \cite{Sato1974a,Noakes1990a},  La$_{2-x}$Sr$_{x}$CuO$_{4}$ \cite{Vignolle2007a} and more recently FeTe$_{1-x}$Se$_{x}$ \cite{Lumsden2009b}.  The Sato-Maki cross-section is of the form:
\begin{equation}
\label{Eq:Sato_Maki} \chi^{\prime\prime}({\mathbf q},\omega)=\chi_\delta(\omega)
\frac{\kappa^4(\omega)} {[\kappa^2(\omega)+R(\mathbf{q})]^2}
\end{equation}
with
\begin{eqnarray*}
R(\mathbf{q}) & = &
\frac{1}{4 \delta^2} \left\{ \left[(h-h_0)^2+(k-k_0)^2-\delta^2 \right]^2 \right. \\
 & + & \left. \frac{\lambda}{4}  \left[(h-h_0)^2- (k-k_0)^2 \right]^2  \right\},
\end{eqnarray*}
where $\mathbf{Q}_{0}=(h_0,k_0)$ is the nearest reciprocal lattice point to $\mathbf{Q}$ with odd $(h+k)$. The location of the peaks in the response is controlled by $\delta$, and its shape by $\kappa$ and $\lambda$. For example, to obtain a peak centered at $\mathbf{Q}$=(1/2,1/2), $\delta=1/\sqrt{2}$. We found that the Sato-Maki function could give a good description of our data except at the highest energies $E \approx 80$~meV. Fits of the SM-function are shown in Figs.~\ref{Fig:slices}-\ref{Fig:cuts}.  The values of $\chi^{\prime\prime}(\omega)$ computed from the SM fits are statistically indistinguishable from those obtained using the SW cross-section.
\section{Discussion}

The magnetic response has been studied by neutron scattering in a number of superconducting ferro-pnictides \cite{Christianson2008a,Lumsden2009a,Chi2009a,Li2009a,Lumsden2009b,Inosov2009a} including Ba(Fe$_{1-x}$Co$_{x}$)$_2$As$_2$ \cite{Lumsden2009a,Inosov2009a}.  Previous experiments on the superconducting 122 systems have either been on powder samples \cite{Christianson2008a} or performed over a smaller energy range \cite{Lumsden2009a,Inosov2009a} than the present experiment.  One of the early results from INS on the iron-based superconductors was the observation of a `spin-resonance' \cite{Christianson2008a,Lumsden2009a,Chi2009a,Inosov2009a}. The spin-resonance is most easily interpreted in terms of a sign difference in the superconducting gap $\Delta(\mathbf{k})$ between different parts of the Fermi surface \cite{Mazin2008a,Barzykin2008a,Korshunov2008a,Cvetkovic2009a,Chubukov2008a,Graser2009a} and is consistent with a $s^{\pm}$ gap function. At the lowest energies probed by the present experiment we are able to observe the spin-resonance. Fig.~\ref{Fig:chi_local} shows this most directly: There is a peak in the local susceptibility $\chi^{\prime\prime}(\omega)$ near 10~meV at $T=7$~K which is suppressed on raising the temperature to 26~K. The resonance is strongest near $\mathbf{q}=(1/2,1/2)$ as shown in Fig.~\ref{Fig:slices}(a) and Fig.~\ref{Fig:cuts}(d,h).
As the energy is increased above the resonance energy, magnetic excitations disperse in an anisotropic manner (see Fig.~\ref{Fig:dispersion}). Thus our results are consistent with an upwardly dispersing mode which is strongest near $\mathbf{q}=(1/2,1/2)$. This is at least qualitatively consistent with theoretical predictions for a $s^{\pm}$ state \cite{Korshunov2008a}. Interestingly, this contrasts with the behavior in YBa$_2$Cu$_3$O$_{6+x}$, where the resonance mode disperses downwards in energy \cite{Mook1998a,Arai1999a,Bourges2000a}.

An important result from our experiment is the observation of dispersive anisotropic spin fluctuations up to 80 meV.  Neutron scattering studies on the antiferromagnetic parents of the 122 series, BaFe$_2$As$_2$ \cite{Ewings2008a,Matan2009a} and CaFe$_2$As$_2$ \cite{Diallo2009a,Zhao2009a}, have observed spin-wave excitations up to about 200~meV.  Unfortunately, single crystal measurements up to 80 meV only exist for CaFe$_2$As$_2$. The magnetic excitations have been analyzed using a spin-wave model based on localized moments.  Ba(Fe$_{0.935}$Co$_{0.065}$)$_2$As$_2$ is close to magnetic order at low temperature. In order to parameterize our data we use a damped spin-wave cross-section with an energy gap. In our case, the gap $\Delta$ in Eq.~\ref{Eq:dispersion} is due to the superconductivity (and the formation of the resonance described above). We find that Eqs.~\ref{Eq:DHO}--\ref{Eq:dispersion} provide a good description of the data (as illustrated by Figs.~\ref{Fig:slices}--\ref{Fig:cuts} ) with $\Gamma/E = 0.15$ (as per Ref.~\onlinecite{Zhao2009a}). This method of analysis produces spin-wave velocities of $v_{\parallel}=580 \pm 60$~meV\AA, $v_{\perp}=230 \pm 30$~meV\AA\ and $\Delta=10 \pm 0.5$~meV.  Using the published exchange constants, we obtain spin-wave velocities for the parent compound CaFe$_2$As$_2$ of $v_{\parallel}$=513~meV\AA\ \cite{Diallo2009a} or 494~meV\AA\ \cite{Zhao2009a} and $v_{\perp}$=370~meV\AA\ \cite{Diallo2009a} or 348~meV\AA\ \cite{Zhao2009a}.  If the magnetic interactions are described using a Heisenberg Hamiltonian
\begin{equation}
H=J_{1}\sum_{<jk>}  \mathbf{S}_j \cdot \mathbf{S}_k +J_{2} \sum_{\ll jk \gg} \mathbf{S}_j \cdot \mathbf{S}_k,
\label{Eq:Heisenberg}
\end{equation}
where the sums are over nearest-neighbor and next-nearest-neighbor pairs (see Fig.~\ref{Fig:unit_cell}),
we can convert the spin-wave velocities to effective exchange constants based on spin-wave theory in the ordered parent antiferromagnets.  Using the relations $v_{\parallel}=S \sqrt{2} a(J_1+2 J_2)$ and $v_{\perp}=S \sqrt{2} a \sqrt{(2 J_2+J_1)(2J_2-J_1)}$ (see Refs.~\onlinecite{Fang2008a,Ewings2008a,Diallo2009a,Zhao2009a}), we obtain $J_1=43 \pm 7$ and $J_2=30 \pm 3$~meV.  It should be noted that this is not the only possible analysis, but it is the simplest.  It may also be possible to interpret the data in terms of the $J_{1a}$-$J_{1b}$-$J_{2}$ model \cite{Yildirim2008a,Zhao2009a} which has been used to describe the ordered parent CaFe$_2$As$_2$. However $|J_{1a}-J_{1b}|$ is certainly less for Ba(Fe$_{0.935}$Co$_{0.065}$)$_2$As$_2$.

In addition to comparing the spin-wave velocities with the parent compounds, we can also compare the strength of the magnetic response. Fig.~\ref{Fig:chi_local} shows the wavevector-averaged or local susceptibility $\chi^{\prime\prime}(\omega)$ defined by Eq.~\ref{Eq:chi_local}.  In the case of Ba(Fe$_{0.935}$Co$_{0.065}$)$_2$As$_2$, we have averaged over $0 \leq h<1$ and $0 \leq k<1$, and $\pm 0.25$ around various odd, even and non-integer $l$ values (see caption to Fig.~\ref{Fig:chi_local}). As mentioned above, points with different $l$ appear to follow the same trend, suggesting that there is no $l$-dependence to this partially averaged quantity.  Thus, the graph represents the true $\chi^{\prime\prime}(\omega)$.  For comparison, we have computed $\chi^{\prime\prime}(\omega)$ for CaFe$_2$As$_2$ using the exchange constants and $S_{\mathrm{eff}}$ from Ref.~\onlinecite{Zhao2009a}. This result is shown as the dashed line \cite{Zhao_comment} in Fig.~\ref{Fig:chi_local}.  We note that the response is slightly larger in Ba(Fe$_{0.935}$Co$_{0.065}$)$_2$As$_2$ over the energy range investigated here.   The same phenomenon occurs in the cuprates where, for example, the spin fluctuations are stronger in optimally-doped La$_{2-x}$Sr$_{x}$CuO$_{4}$ than in La$_{2}$CuO$_{4}$ \cite{Hayden1996a,Vignolle2007a}. One might understand this increase at lower energies as being due to a shift in spectral weight from higher energies (above the window of the present experiment) and the loss of the peak due to magnetic order present in the ordered compounds.

It is interesting to compare the present results with those obtained on the cuprates.  Firstly, we note that the magnitude of the local susceptibility is similar in optimally doped La$_{2-x}$Sr$_{x}$CuO$_{4}$ \cite{Hayden1996a,Vignolle2007a} and Ba(Fe$_{0.935}$Co$_{0.065}$)$_2$As$_2$.  Both the cuprates and Ba(Fe$_{0.935}$Co$_{0.065}$)$_2$As$_2$ show a strong response near the $\mathbf{Q}$=(1/2,1/2). In La$_{2-x}$Sr$_{x}$CuO$_{4}$ and YBa$_2$Cu$_3$O$_{6+x}$, various dispersive modes and `resonance' features \cite{Mook1998a,Arai1999a,Bourges2000a,Vignolle2007a} are observed below $E \lesssim 50$~meV. At higher energies, $50 \lesssim E \lesssim 100$~meV, the response in YBa$_2$Cu$_3$O$_{6+x}$ and La$_{2-x}$Sr$_{x}$CuO$_{4}$ is quasi-isotropic or has fourfold symmetry about the (1/2,1/2) position \cite{Hayden2004a,Stock2005a,Hinkov2007a,Vignolle2007a}. In contrast, the response in Ba(Fe$_{0.935}$Co$_{0.065}$)$_2$As$_2$ is more anisotropic being broader along $(1/2-\xi,1/2+\xi)$  than the $(1/2+\xi,1/2+\xi)$. We discuss possible origins of this anisotropy below.

Discussion about the best way to describe the magnetic interactions in the ferropnictides continues \cite{Yildirim2008a,Fang2008a,Singh2009a,Johannes2009a}.  These materials are not Mott insulators with localized spins which can be described purely by near-neighbor superexchange.  Neither are the ferropnictides weakly correlated metallic systems.  It appears that Hund's rule coupling is responsible for producing the atomic moments and there is also strong magnetoelastic coupling. Keeping this is mind, we discuss our results in terms an effective Heisenberg coupling and a weakly correlated metallic picture.

There is evidence that the ferropnictides develop an underlying electron nematic phase \cite{Fang2008a,Xu2008a,Mazin2009a} as they approach the magnetic ordering temperature.  That is, they show fluctuating magnetic stripes based on the ordering shown in Fig.~\ref{Fig:unit_cell}(a).  While Ba(Fe$_{0.935}$Co$_{0.065}$)$_2$As$_2$ does not appear to order magnetically, it is close to the antiferromagnetic quantum critical point \cite{Ni2008a,Chu2009a,Lester2009a}, therefore similar considerations should apply.  It has been known for some time \cite{Chandra1990a} that 2D square-lattice systems described by the Hamiltonian in Eq.~\ref{Eq:Heisenberg} develop a stripe-like order of the type observed in the ferropnictides for
$J_{2}/J_{1} > 1/2$ which is the condition here (our measurements give
$J_{2}/J_{1}=0.70 \pm 0.09$).  Although the Hamiltonian in Eq.~\ref{Eq:Heisenberg} has a $C_4$ lattice symmetry, we expect low-frequency correlations to develop at low temperature through the so-called `order out of disorder' mechanism \cite{Chandra1990a,Fang2008a,Xu2008b}.  This means instantaneous collinear order described by two interpenetrating N\'{e}el sublattices as shown in Fig.~\ref{Fig:unit_cell}(a). The higher-energy magnetic excitations of such a state with long-lived correlations of this type might be expected to be similar to the excitations of the corresponding ordered antiferromagnet. This could explain the similarities between the magnetic excitations in the $10<E<80$~ meV range in Ba(Fe$_{0.935}$Co$_{0.065}$)$_2$As$_2$ and CaFe$_2$As$_2$ \cite{Diallo2009a,Zhao2009a}.

The second approach which has been used to describe the ferropnictides is that based on nesting between the electron and hole pockets \cite{Korshunov2008a,Chubukov2008a,Graser2009a,Yaresko2009a,Singh2009a}. In such a picture, the magnetic excitations are correlated electron-hole pairs.  The canonical example of an itinerant antiferromagnet which is described by such a picture is elemental chromium \cite{Fawcett1988a}.  Model Lindhard calculations of the wavevector-dependent susceptibility $\chi(\mathbf{q},\omega)$ \cite{Graser2009a,Yaresko2009a} based on the band structure appear to reproduce the $\mathbf{q}$-anisotropy in $\chi^{\prime\prime}(\mathbf{q},\omega)$ observed here at higher energies. In particular, the response is broader along $(1/2-\xi,1/2+\xi)$ rather than the $(1/2+\xi,1/2+\xi)$ for a given energy. It is interesting to note that highly structured magnetic excitations, characteristic of nesting, have recently been observed in FeTe$_{1-x}$Se$_{x}$ \cite{Qiu2009a,Lumsden2009b,Li2010a}. For FeTe$_{0.51}$Se$_{0.49}$, an anisotropic response similar to the one reported here is observed. The authors of Ref.~\onlinecite{Lumsden2009b} find that the Sato-Maki cross section (Eq.~\ref{Eq:Sato_Maki}) provides a good description of their data over a 10--120~meV energy range.   Motivated by this we also fitted our data to the Sato-Maki form (see Fig.~\ref{Fig:slices}). While the Sato-Maki function provides a reasonable description of the data, the phenological spin-wave cross section provides a better description at higher energies.

\section{Conclusions}

In summary, we have used inelastic neutron scattering  to probe the collective spin excitations of near optimally doped  Ba(Fe$_{1-x}$Co$_{x}$)$_2$As$_2$ ($x$=0.065).  Strongly dispersive spin fluctuations are observed up to 80~meV.  In the superconducting state, our measurements are consistent with a mode dispersing from the spin resonance near $\mathbf{Q}=(1/2,1/2)$ to high energy.  At higher energies, we observe excitations which are anisotropic as a function of in-plane wavevector.  The response is centered on the $M$ or (1/2,1/2) position of the Brillouin zone and is
broader along $(1/2-\xi,1/2+\xi)$ rather than the $(1/2+\xi,1/2+\xi)$ direction.  This anisotropy may be understood in terms of multiple $J_{1}$-$J_{2}$ exchange interactions or response functions derived from the band structure.   When placed on an absolute scale, our measurements show that the local- or wavevector-integrated susceptibility is generally larger in magnitude than that of the ordered parent antiferromagnets over the energy range probed here.

\textit{Note Added.} Diallo et al. \cite{Diallo2010a} have recently reported magnetic fluctuations in the paramagnetic state of CaFe$_2$As$_2$ which are similar to those presented here.

\section*{Acknowledgements}
We acknowledge useful discussions with J.P. Rodriguez, Pengcheng Dai and Jun Zhao.


\begin{thebibliography}{10}
\bibitem{chi_comment}
In the present experiment we assume $\chi^{\prime\prime}_{\alpha
  \alpha}(\mathbf{q},\omega)$ is isotropic with respect to field direction.
  Thus the susceptibilities quoted correspond to
  $\chi^{\prime\prime}$=$(1/3)(\chi^{\prime\prime}_{xx}+\chi^{\prime\prime}_{y%
y}+\chi^{\prime\prime}_{zz}$).
\end{thebibliography}

\end{document}